\documentclass{PoS}
\newcommand{\msbar}{\overline{\footnotesize\textrm{{MS}}}}

\title{Radiative corrections to decay amplitudes in lattice QCD}

\ShortTitle{Radiative corrections}

\author{Davide Giusti$^{(a)(b)}$, 
Vittorio Lubicz$^{(a)(b)}$, Guido Martinelli$^{(c)}$,
\speaker{Christopher~Sachrajda}\nolinebreak$^{\,(d)}$,
Francesco Sanfilippo$^{(b)}$, Silvano Simula$^{(b)}$, \hspace{1in} Nazario Tantalo$^{(e)}$
\\$^{(a)}$ Dip. di Matematica e Fisica, Universit\`a Roma Tre, Via della Vasca Navale 84, I-00146 Rome, 
\nolinebreak Italy\\ 
$^{(b)}$ INFN, Sezione di Roma Tre, Via della Vasca Navale 84, I-00146 Rome, Italy\\
$^{(c)}$ Dipartimento di Fisica, Universit\`a di Roma La Sapienza and INFN Sezione di Roma, Piazzale Aldo Moro 5, 00185 Roma, Italy\\
        $^{(d)}$ Dept. of Physics and Astronomy, University of Southampton, Southampton SO17 1BJ, UK\\
$^{(e)}$ Dip. di Fisica, Universita` di Roma ``Tor Vergata" and INFN Sezione di Tor Vergata, \\ Via della Ricerca Scientifica 1, I-00133 Roma, Italy\\ 
        E-mail: \email{davide.giusti@uniroma3.it}, \email{lubicz@fis.uniroma3.it}, 
        \email{guido.martinelli@roma1.infn.it}, \email{cts@soton.ac.uk}, \email{francesco.sanfilippo@infn.it},
        \email{silvano.simula@roma3.infn.it}, \email{nazario.tantalo@roma2.infn.it}}

\abstract{The precision of lattice QCD computations of many quantities has reached such a precision that isospin-breaking corrections, including electromagnetism, must be included if further progress is to be made in extracting fundamental information, such as the values of Cabibbo-Kobayashi-Maskawa matrix elements, from experimental measurements.
We discuss the framework for including radiative corrections in leptonic and semileptonic decays of hadrons, including the treatment of infrared divergences. We briefly review isospin breaking in leptonic decays and present the first numerical results for the ratio $\Gamma(K_{\mu2})/\Gamma(\pi_{\mu2})$ in which these corrections have been included. We also discuss the additional theoretical issues which arise when including electromagnetic corrections to semileptonic decays, such as $K_{\ell3}$ decays. The separate definition of strong isospin-breaking effects and those due to electromagnetism requires a convention. We define and advocate conventions based on hadronic schemes, in which a chosen set of hadronic quantities, hadronic masses for example, are set equal in QCD and in QCD+QED.  This is in contrast with schemes which have been largely used to date, in which the renormalised $\alpha_s(\mu)$ and quark masses are set equal in QCD and in QCD+QED in some renormalisation scheme and at some scale $\mu$.}

\FullConference{The 36th Annual International Symposium on Lattice Field Theory - LATTICE2018\\
		22-28 July, 2018\\
		Michigan State University, East Lansing, Michigan, USA.}

\begin{document}

\vspace{-0.1in}\section{Introduction} 

\vspace{-0.1in}The hugely impressive recent progress in lattice QCD computations has led to the determination of many quantities, including leptonic decay constants and semileptonic form factors, of $O(1\%)$ or better\,\cite{Aoki:2016frl}. In order to make further progress in determining the CKM matrix elements from these decays, isospin-breaking effects, including radiative corrections, must be included. This requires a procedure for handling the presence of infrared divergences, as well as an understanding of the finite-volume effects due to the presence of the photon. In this talk we review the status of our proposal for how this might be achieved~\cite{Carrasco:2015xwa,
Lubicz:2016xro} and present the first numerical results for the ratio of leptonic decay rates $\Gamma(K_{\mu2})/\Gamma(\pi_{\mu2})$~\cite{Giusti:2017dwk}. Ongoing work in extending the framework to semileptonic decays is reviewed with a discussion of the new theoretical issues which arise in this case. We start however, with a discussion of the convention-dependent question of how one might separate strong isospin breaking effects from those due to radiative corrections.

\vspace{-0.1in}\section{What is QCD?} \label{sec:whatisQCD} 

\vspace{-0.1in}In the \emph{full} (QCD+QED) theory what one means by QCD and what one means by radiative corrections becomes convention dependent. To illustrate this consider the action for the full theory:
\begin{equation}
S^{\mathrm{\,full}}=\frac{1}{g_s^2}S^{\mathrm{YM}}+\sum_f\left\{S_f^{\mathrm{kin}}+m_fS_f^{\mathrm{m}}\right\}+S^{\mathrm A}+\sum_\ell \left\{S^{\mathrm{kin}}_\ell+m_\ell S_\ell^{\mathrm{m}}\right\}\,,
\end{equation}
where $S^{\mathrm{YM}}$ and $S^\mathrm{A}$ are the gluon and Maxwell actions respectively and the sums are over the 
flavours of the quarks $f$ and the charged leptons $\ell$, with ``$\textrm{kin}$" and "$\textrm{m}$" labelling the corresponding kinetic and mass terms. We imagine calculating some observable $O$ with this action and this is unambiguous. At the level of $O(1\%)$ however, what the QCD and QED contributions to $O$ are separately requires a definition. Before explaining this, it may be instructive to recall how the choice of bare quark masses $m_f$ and the strong coupling $g_s$ is made in the absence of QED.  

\vspace{0.1in}\noindent\textbf{Calculation of O in the absence of QED:}
When performing QCD simulations in the 4-flavour theory without QED, for each value of $g_s$ we can, for example, choose the four \emph{physical} bare quark masses ($m_u^0, m_d^0, m_s^0, m_c^0$) to be those for which the 4 dimensionless ratios:
\begin{equation}\label{eq:Ridef}
R_1=\frac{a_0m_{\pi^0}}{a_0m_\Omega},~R_2=\frac{a_0m_{K^0}}{a_0m_\Omega},~R_3=\frac{a_0m_{K^+}}{a_0m_\Omega}~\textrm{and}~R_4=\frac{a_0m_{D^0}}{a_0m_\Omega},
\end{equation}
take their physical values. The superscript $0$ on the bare masses and the subscript $0$ on the lattice spacing $a_0$ below indicate that these quantities are defined in QCD, i.e. without QED.
The lattice spacing $a_0$ corresponding to the chosen value of $g_s$ is obtained by imposing that some dimensionful quantity, e.g. the mass of the $\Omega$-baryon, takes its physical value:
\begin{equation}\label{eq:a0def}
a_0=\frac{a_0m_\Omega}{m_\Omega^{\scriptsize\textrm{phys}}}\,.
\end{equation}
Of course different choices for the quantities used to determine the $m_{u,d,s,c}^0$ and $a_0$ can be made, leading to different lattice artefacts in predictions. Once QED corrections are included however, the masses of hadrons $H$ are shifted by $O(\alpha)m_H$ and so some choice of convention is necessary if we wish to define the QCD and QED contributions separately.

\vspace{0.1in}\noindent\textbf{Hadronic scheme(s):}
With the inclusion of QED, in the hadronic scheme (\underline{which we advocate}) we define QCD by imposing exactly the same conditions as above for QCD without QED. The QED corrections now shift the 
hadronic masses used for the calibration and to compensate for this we add mass counterterms 
$m_f=m_f^0+\delta m_f$ 
so that the ratios $R_i$ in Eq.(\ref{eq:Ridef}) take their physical values. The $\Omega$ mass itself is 
changed, leading to a shift in the lattice spacing, $a=a_0+\delta a$.

Having calibrated the lattice, imagine we wish to make a prediction for an observable $O$, which for illustration we take to be of mass dimension 1:
\begin{equation}
O=\frac{\langle aO\rangle^{\scriptsize\textrm{full}}}{a}=
\frac{\langle a_0O\rangle^{\scriptsize\textrm{QCD}}}{a_0}+\frac{\delta O}{a_0}-\frac{\delta a}{a_0^2} \langle a_0O\rangle^{\scriptsize\textrm{QCD}}~+\,O(\alpha^2)\,,
\end{equation}
where $\delta O$ is the contribution from the electromagnetic corrections and mass counterterms.
The first term on the right-hand side can be calculated within QCD alone and has a well defined continuum limit as does the sum. Such a separation allows us to answer the question: ``What is the difference between QCD (defined as above) and the full theory"? The isospin breaking corrections are calculated directly, i.e. without taking the difference between calculations performed in the full theory and in QCD. If needed in the future, the scheme can be extended to higher orders in $\alpha$.

\vspace{0.1in}\noindent\textbf{The GRS scheme:}
Other ways of defining what the QED corrections are clearly possible. An indirect one has been proposed by Gasser, Rusetsky and Scimemi~\cite{Gasser:2003hk}, and has been followed in a number of publications, including the reviews of the \emph{Flavour Physics Lattice Averaging Group} (FLAG)\,\cite{Aoki:2016frl}. In this scheme the renormalised coupling and masses are equal in QCD and QCD+QED in some scheme and at some renormalisation scale
\begin{eqnarray}
g_s(\mu)&=&Z_g(0,g_s^0,\mu)\,g_s^0=Z_g(e,g_s,\mu)\,g_s\nonumber\\
m_f(\mu)&=&Z_{m_f}(0,g_s^0,\mu)\,m_{f}^{0}(0,g_s^0)=Z_{m_f}(e,g_s,\mu)\,m_{f}(e,g_s)\,,
\end{eqnarray}
where $g_s^0$ is the bare coupling in QCD.
FLAG has adopted such a definition in the $\msbar$ scheme at a scale of $\mu=2\,$GeV.
The four dimensionless ratios $R_i$ ($i=1$-4) in Eq.(\ref{eq:Ridef})
no longer take their physical values and we can write:
$R_i=R_i^{\scriptsize\textrm{phys}}(1+\epsilon_i)\,,$
where the $\epsilon_i$ are $O(\alpha)$.

Since hadronic masses are now calculated precisely in lattice simulations and their values are well known from experimental measurements, we strongly suggest that it is more natural to use hadronic schemes to define what is meant by QCD. By contrast, the renormalised couplings and masses are derived quantities which are not measured directly in experiments.

\vspace{-0.1in}\section{Radiative corrections in leptonic decays}\label{sec:leptonic}
\begin{figure}[t]
\begin{center}
\includegraphics[width=0.8\hsize]{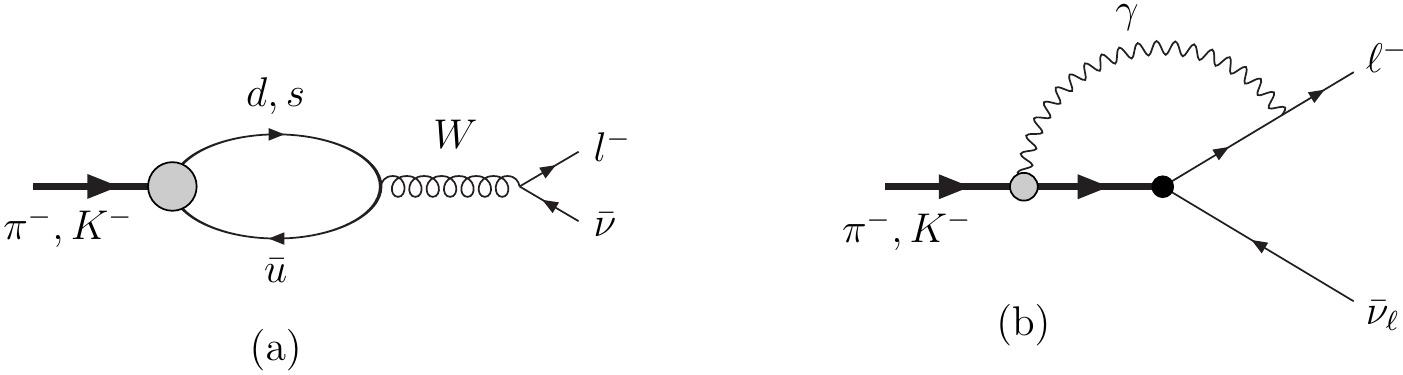}\hspace{0.5in}
\caption{(a) Schematic diagram illustrating the leptonic decays of $\pi$ and $K$ mesons in QCD. (b) one of the  diagrams contributing to the radiative corrections to $K_{\ell 2}$ decays; the solid circle represents the insertion of the effective weak Hamiltonian. \label{fig:Pl2}}
\end{center}
\end{figure}

\vspace{-0.1in}We now consider leptonic decays of pseudoscalar mesons illustrated for $K_{\ell 2}$ and $\pi_{\ell 2}$ decays in Fig.\,\ref{fig:Pl2}(a).  Without QED corrections all the QCD effects are contained in a single number, e.g. for $K_{\ell 2}$ decays it is the kaon decay constant $f_K$ ($\langle
0|\,\bar s\gamma^\mu\gamma^5 u\,|K(p)\rangle\equiv if_K\,p^\mu$), and the rate is given by
\begin{equation}\label{eq:Kl2}
\Gamma(K^-\to\ell^-\bar\nu_{\ell})=\frac{G_F^2\,|V_{us}|^2f_K^2}{8\pi}\,m_K\,m_\ell^2\left(1-\frac{m_\ell^2}{m_K^2}\right)^{\!\!\!2}\,,
\end{equation}
where $G_F$ is the Fermi constant.
Equation (\ref{eq:Kl2}) is a particularly simple paradigm illustrating the approach taken in much of precision flavour physics. On the left-hand side is a quantity which is measured experimentally. On the right-hand side is the fundamental quantity to be determined (in this case $V_{us}$) and also term(s) containing the non-perturbative QCD effects (here $f_K$) and known kinematic factors. Thus once $f_K$ is determined we can use (\ref{eq:Kl2}) to obtain $V_{us}$ (although in practice it is more useful to compare the rates for $K_{\ell 2}$ and $\pi_{\ell 2}$ decays and obtain the ratio $V_{us}$/$V_{ud}$\,\cite{Aoki:2016frl}).

If we require a precision of better than about 1\%, then radiative corrections must be included and one contributing diagram  is sketched in Fig.\ref{fig:Pl2}(b). From this diagram we can see that the strong and electroweak interactions no longer factorise and that the non-perturbative strong interactions are no longer  contained in the single number $f_K$. Moreover, diagrams such as these are infrared divergent, as can be seen by simple power counting; in the region of small photon momenta $k$, the photon propagator $\sim 1/k^2$ and the meson and lepton propagators each $\sim 1/k$, leading to an integrand which behaves as $1/k^4$. In order to construct a physical observable which is free of infrared divergences we must include real photons in the final state\,\cite{Bloch:1937pw}; indeed any measurement of the leptonic decay rate will necessarily include the contributions of real photons with energies below the resolution of the detector. We have proposed a method for computing radiative corrections to decay rates in lattice simulations, including the handling of infrared divergences, as we now explain.  

%\subsection{Lattice computations of {\boldmath$\Gamma(K^+\to\ell^+\nu_{\ell}(\gamma))$} at {\boldmath$O(\alpha)$}}

At this stage, the observable we calculate is $\Gamma_0(K\to\ell\bar\nu_\ell)+\Gamma_1(K\to\ell\bar\nu_\ell\gamma)$, where the subscripts $0,1$ denote the number of photons in the final state and the energy of the photon in the kaon rest-frame $E_\gamma$, satisfies $E_\gamma<\Delta E$ where the cut-off $\Delta E$ is sufficiently small for the structure dependence of $K$ to be neglected
\footnote{We are also developing techniques to compute $\Gamma_1$ in lattice simulations, removing the need for a low value of $\Delta E$.}. 
In practice one might take $\Delta E\lesssim 20$\,MeV which is possible with the energy resolution of the KLOE\,\cite{Ambrosino:2005fw,Ambrosino:2009aa} and NA62 experiments. It is now very useful to rewrite the observable in the form
\begin{equation}
\Gamma_0+\Gamma_1(\Delta E)=\lim_{V\to\infty}(\Gamma_0-\Gamma_0^{\mathrm{pt}})+
\lim_{V\to\infty}(\Gamma_0^{\mathrm{pt}}+\Gamma_1(\Delta E))\,,
\end{equation}
where $\Gamma_0^{\mathrm{pt}}$ is the width calculated for a \emph{point-like} kaon and $V$ is the volume. The second term on the right-hand side can be calculated in perturbation theory directly in infinite-volume. It is infrared convergent, but contains a term proportional to $\log\Delta E$; the result is presented in Ref.\,\cite{Carrasco:2015xwa}.
The first term is also infrared convergent, since when the virtual photon in diagrams for $\Gamma_0$, see Fig.\,\ref{fig:Pl2}(b) for example, is soft it couples only to the charge of the meson.  $\Gamma_0^\mathrm{pt}$ is calculated in perturbation theory whereas $\Gamma_0$ itself must be computed non-perturbatively in a lattice simulation since hard modes, which do resolve the structure of the kaon, contribute to diagrams such as that in Fig.\,\ref{fig:Pl2}(b).

We use QED$_\mathrm{L}$ to regulate the zero-mode in the photon propagator in a finite volume\,\cite{Hayakawa:2008an}. The finite-volume effects then take the form:
\begin{equation}
\Gamma_0^{\mathrm{pt}}(L) =  C_0(r_\ell) + \tilde C_0(r_\ell)\log\left(m_\pi L\right)+ \frac{C_1(r_\ell)}{m_\pi L}+ 
\dots \,, \end{equation}
where $r_\ell=m_\ell/m_K$ and $m_\ell$ is the mass of the final-state charged lepton $\ell$. The coefficients $C_0(r_\ell)$, $\tilde C_0(r_\ell)$ and $C_1(r_\ell)$ are \emph{universal}, i.e. independent of the structure of the meson beyond the value of $f_K$ computed in QCD, so that the leading structure-dependent finite-volume effects in $\Gamma_0-\Gamma_0^{\mathrm{pt}}$ are of $O(1/L^2).$ We present the explicit expressions for $C_{0,1}(r_\ell)$ and $\tilde C_0(r_\ell)$ in Ref.\,\cite{Lubicz:2016xro}.

As a first numerical study of the method, we have computed the ratio $\Gamma(K_{\mu 2})/\Gamma(\pi_{\mu 2})$ using the twisted mass formulation of lattice fermions. We define the correction $\delta R_{K\pi}$ using the relation
\begin{equation}
\frac{\Gamma(K_{\mu 2})}{\Gamma(\pi_{\mu 2})}=\left|\frac{V_{us}}{V_{ud}}\frac{f_K^{(0)}}{f_\pi^{(0)}}\right|^2\,\frac{m_\pi^3}{m_K^3}\,\left(\frac{m_K^2-m_\mu^2}
{m_\pi^2-m_\mu^2}\right)^{\!\!2}\,\left(1+\delta R_{K\pi}\right)\,,
\end{equation}
where $m_{K,\pi}$ are the physical masses and (in spite of the discussion in Sec.\ref{sec:whatisQCD}) $f_{K,\pi}^{(0)}$ are the decay constants obtained in iso-symmetric QCD with the renormalized $\msbar$ masses and coupling equal to those in the full QCD+QED theory extrapolated to infinite volume and to the continuum limit\,\cite{Giusti:2017dwk}.
Using numerous twisted-mass ensembles we find
$\delta R_{K\pi}=-0.0122(16)\,.$ With this definition of the correction we can compare directly with the same quantity obtained in chiral perturbation theory $\delta R_{K\pi}=-0.0112(21)$\,\cite{Cirigliano:2011tm}, where the unknown low-energy constants cancel in the ratio.

\begin{figure}[t]
\begin{center}
\includegraphics[width=0.48\hsize]{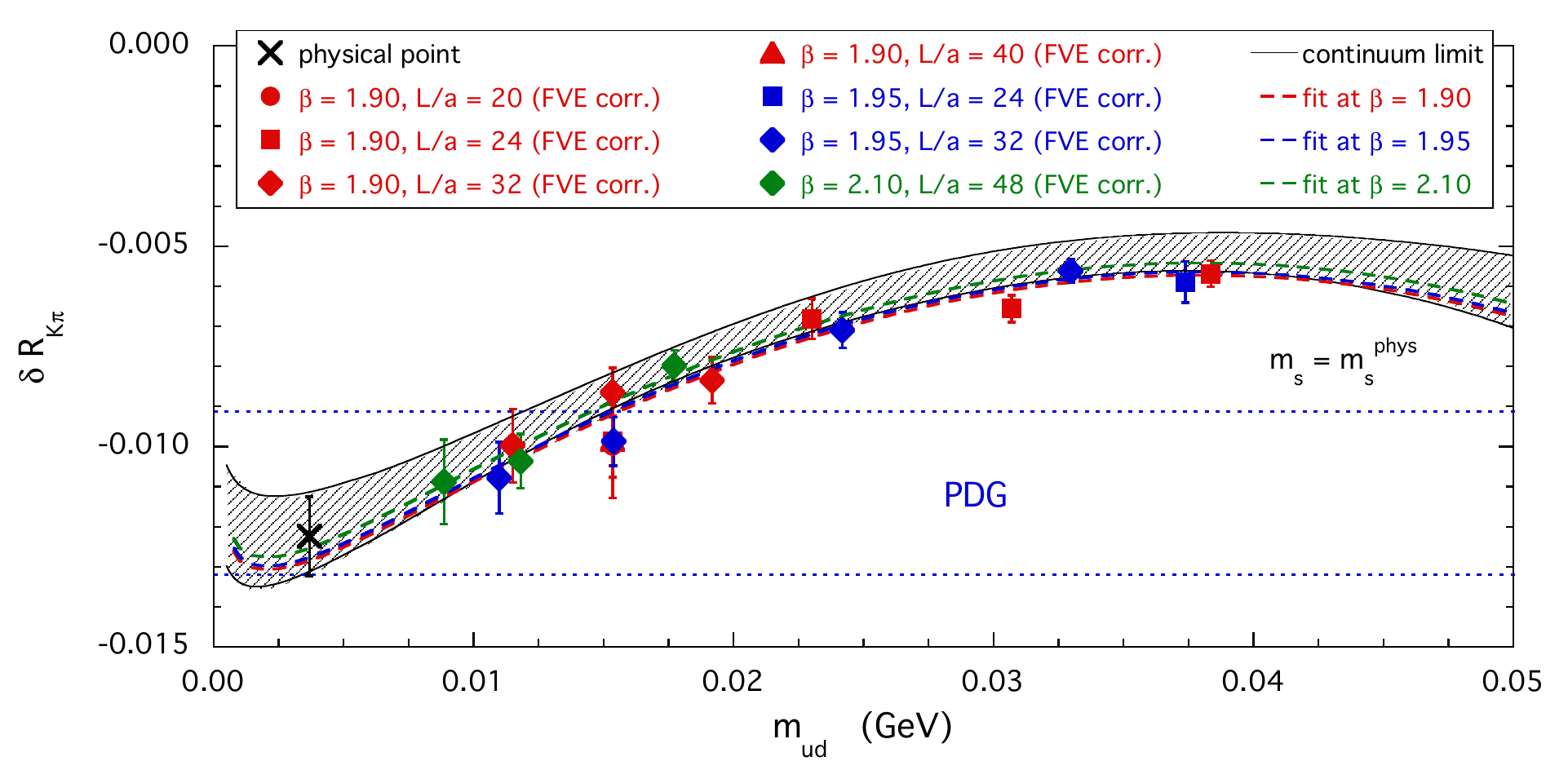}\hspace{0.2in}
\includegraphics[width=0.48\hsize]{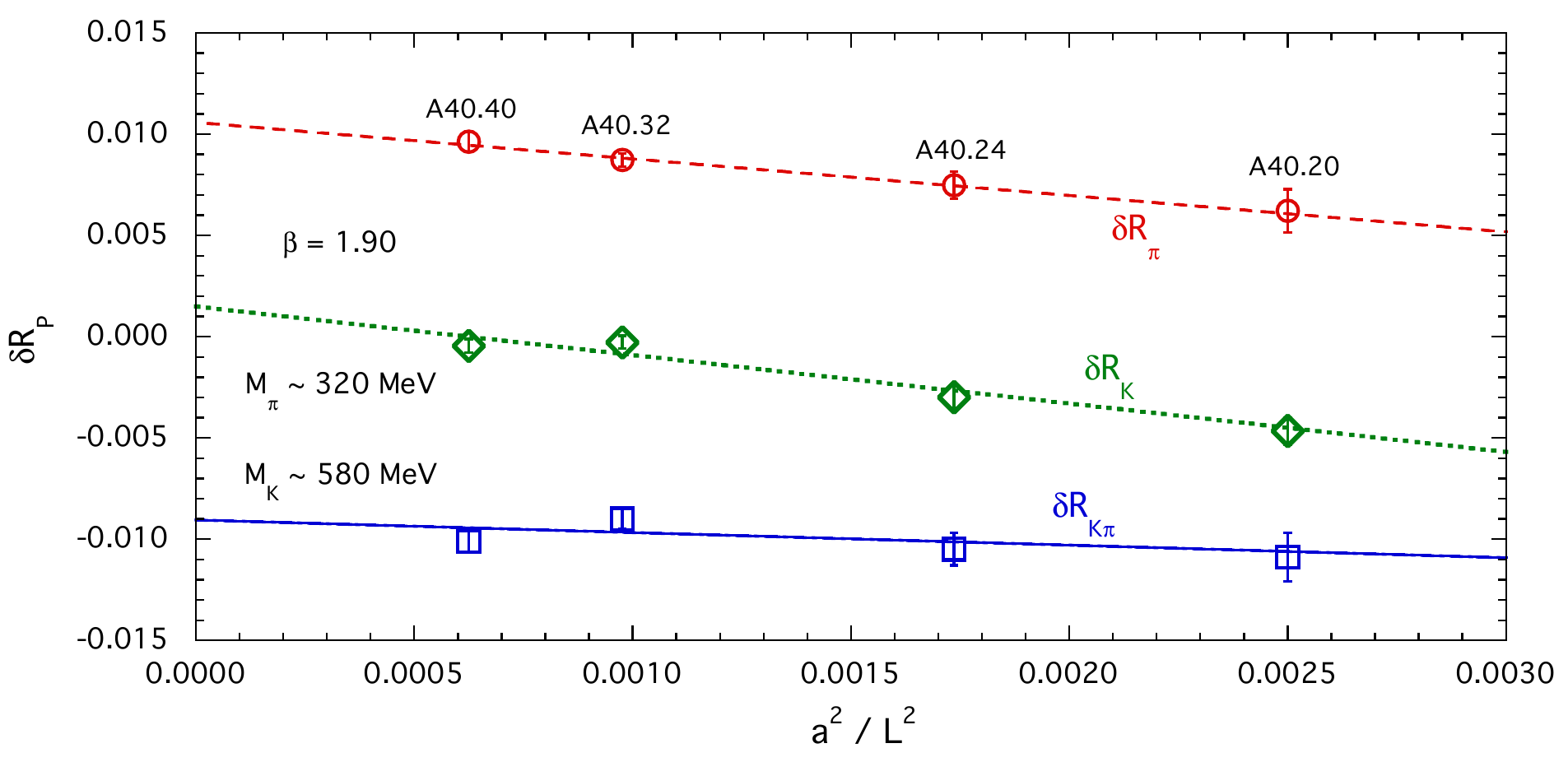}
\end{center}
\caption{Left-hand plot: results for the correction $\delta R_{K\pi}$ obtained using a range of quark masses, $\beta$s and volumes with the twisted-mass formulation for the fermions. The cross corresponds to the extrapolation to the physical point. Right-hand plot: Volume dependence of results for $\delta R_{K,\pi}$, and for $\delta R_{K}$ and $\delta R_{\pi}$ separately, obtained at the same $\beta$ and meson masses (as indicated) after the universal terms proportional to $\log[m_{\pi,K}L]$ and $1/L$ have been subtracted.\label{fig:numerical}}
\end{figure}

In Fig.\,\ref{fig:numerical} we present two plots from \cite{Giusti:2017dwk} showing the results from the individual ensembles as well as the extrapolation to the physical point (left-hand plot) and the volume dependence (after subtraction of the universal terms) of results obtained at four different volumes at the same masses and $\beta$.  The calculation has been performed in the electro-quenched approximation. Full details, and the separate results for $\Gamma(K_{\mu 2})$ and $\Gamma(\pi_{\mu 2})$ will be presented in a paper currently in preparation.

\vspace{-0.1in}\section{Radiative corrections in semileptonic decays}\label{sec:semileptonic}

\vspace{-0.1in}We are now generalising the framework developed for leptonic decays and described in Sec.\,\ref{sec:leptonic} to semileptonic decays, such as $\bar{K}^0\to\pi^+\ell^-\bar\nu_\ell$ which we will use for illustration. Although the main ideas presented above are also applicable in this case, several new features arise which we now explain.

\begin{figure}[t]
\begin{center}
\includegraphics[width=0.3\hsize]{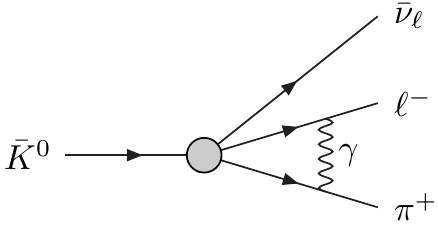}
\end{center}
\caption{One of the diagrams contributing to radiative corrections to $K_{\ell3}$ decays.
\label{fig:semileptdiagram}}
\end{figure}

In QCD the amplitude depends on two form factors $f_{0,+}(q^2)$, where 
%$q$ is the momentum transfer between the $\bar{K}^0$ and the $\pi^+$; 
$q=p_K-p_\pi=p_\ell+p_\nu$:
\begin{equation}
\langle\,\pi(p_\pi)\,|\bar{s}\gamma_\mu u\,|\,K(p_K)\,\rangle=
f_0(q^2)\,\frac{M_K^2-M_\pi^2}{q^2}q_\mu +
f_+(q^2)\,\left[(p_\pi+p_K)_\mu-\frac{M_K^2-M_\pi^2}{q^2}q_\mu
\right]\,.
\end{equation}
When including radiative corrections, the natural observable to consider is $d^2\Gamma/dq^2ds_{\pi\ell}$, where $s_{\pi\ell}=(p_\pi+p_\ell)^2$ and we follow the 
same procedure as for leptonic decays and write:
\begin{equation}
\frac{d^2\Gamma}{dq^2 ds_{\pi\ell}}=\lim_{V\to\infty}
\left(
\frac{d^2\Gamma_0}{dq^2 ds_{\pi\ell}}
-\frac{d^2\Gamma_0^{\scriptsize\textrm{pt}}}{dq^2 ds_{\pi\ell}}\right)
+\lim_{V\to\infty}\left(
\frac{d^2\Gamma_0^{\scriptsize\textrm{pt}}}{dq^2 ds_{\pi\ell}}+\frac{d^2\Gamma_1(\Delta E)}{dq^2 ds_{\pi\ell}}\right)
\end{equation}
where $\Delta E$ is the cut-off on the energy of the real photon. The infrared divergences cancel separately in each of the two terms on the right-hand side. %The second term has been calculated in infinite volume in the eikonal approximation:  $(p-k)^2-m^2\to -2p\cdot k$, where $p$ is the momentum of an external on-shell particle and $k$ is the momentum of the photon~\cite{Isidori:2007zt,deBoer:2018ipi}.

We have determined the integrands/summands which need to be input into the Poisson summation formula to determine the finite-volume corrections. An important difference with leptonic decays is that the $1/L$ corrections depend on the derivatives of the form factors, $df_{0,+}/dq^2$, and not just on the form factors themselves. This is a general feature for amplitudes which depend on external kinematic variables other than hadron masses. The derivatives $df_{0,+}/dq^2$ are physical quantities and in principle can be determined from experiment or lattice calculations. On the other hand the $1/L$ corrections do not depend on derivatives of the form factors with respect to the meson masses; such derivatives are not physical. The explicit calculation of the $1/L$ corrections has not yet been performed. 

A second significant difference with leptonic decays is the presence of intermediate states in the diagram of Fig.\ref{fig:semileptdiagram} which have a lower energy than that of the external pion-lepton pair. The unphysical contributions to the correlation function from such states grow exponentially with the temporal separation of the weak Hamiltonian and the operators at the sink which annihilate the pion and lepton. The presence of such contributions and the need to subtract them is a general feature in the calculation of long-distance effects and is a manifestation of the Maiani-Testa theorem\,\cite{Maiani:1990ca}. The number of such states depends on the choice of the kinematic variables $q^2$ and $s_{\pi\ell}$ as well as on the volume. 
The problem of intermediate states with energies smaller than that of the external state, including those containing additional hadrons, is particularly severe for semileptonic decays of heavy mesons. For much of the phase-space there appear to be too many lighter intermediate states to handle effectively. This is analogous to the fact that the amplitudes and CP-asymmetries for charmless two-body $B$-decays, such as $B\to\pi\pi$ and $B\to\pi K$, are not calculable in lattice calculations, whereas $K\to\pi\pi$ decay amplitudes can be calculated.

\vspace{-0.1in}\section{Future prospects}\label{sec:concs}

\vspace{-0.1in}We are successfully developing and implementing a framework for the \emph{ab initio} calculation of 
radiative corrections to leptonic and semileptonic decays. Such a framework is \underline{necessary} if 
we are to determine the CKM matrix elements to a precision of better that 1\% or so.
The priority for the improvement of the calculation(s) is the renormalization. The effective Hamiltonian for these decays is $H_{\mathrm{eff}}=\frac{G_F}{\sqrt{2}}V_{ij}^{\mathrm{CKM}}\left(1+\frac{\alpha}{\pi}\log\frac{M_Z}{M_W}\right)O_{1}^{\mathrm{W}}$, where for kaon decays $O_{1}^{\mathrm{W}}=(\bar{u}\gamma^\mu_Ls)(\bar{\ell}\gamma_{\mu\,L}\nu_\ell)$ and the superscript $W$ indicates that the operator is to be evaluated in the $W$-regularisation in which the photon propagator is $M_W^2/(k^2(M_W^2-k^2))$. Up to now we have matched the operators in the lattice and W-regularisations at $O(\alpha)$ and are currently in the process of extending this to include the strong interactions.

For semileptonic decays, in addition to the renormalization, some important practical issues need to be investigated, including the subtraction of the unphysical (exponentially growing in time) contributions in an actual computation. 
This requires a phenomenological understanding of the phase-space needed to obtain precise determinations of the CKM matrix elements; what useful cuts can be imposed on $q^2$ and $s_{\pi\ell}$ to facilitate the subtraction of the unphysical intermediate states?

\vspace{0.1in}\noindent\textbf{Acknowledgements} 
%We gratefully acknowledge the CPU time provided by PRACE under Project No. Pra10-2693 and by CINECA under the initiative INFN-LQCD123 on the BG/Q system Fermi. 
V.L., G.M., S.S. and C.T. thank MIUR (Italy) for partial support under Contract No. PRIN 2015P5SBHT. G.M. also acknowledges partial support from ERC Ideas Advanced Grant No. 267985 ``DaMeSyFla". C.S. was partially supported by STFC (UK) Grants No.\,ST/ L000296/1 and No. ST/P000711/1 and by an Emeritus Fellowship from the Leverhulme Trust.

\end{document}